\documentclass[preprint,5p,twocolumn,showpacs,showkeys,10pt] {elsarticle}
\usepackage{color}
\usepackage{graphicx,amssymb,epsf,amsmath} 
\usepackage{epsfig}
\usepackage{epstopdf}

\newcommand{\be}{\begin{equation}}
\newcommand{\ee}{\end{equation}}
\newcommand{\bea}{\begin{eqnarray}}
\newcommand{\eea}{\end{eqnarray}}

\journal{Physics Letters A}

\begin{document}

\begin{frontmatter}

\title{ Application of conditional shape invariance symmetry to obtain the eigen-spectrum of the mixed potential
$ V(r)=ar+br^{2}+ \frac{c}{r}+\frac{l(l+1)}{r^2}   $ }
\author{Sudesna Bera$^{1}$,Barnali Chakrabarti$^{1}$, T.K. Das$^{2}$ }
\address{$^{1}$Department of Physics, Presidency University, 86/1 College Street, Kolkata 700083,}
\address{$^{2}$ Department of Physics, Calcutta University, 92 A.P.C Road, Kolkata-700009}

\begin{abstract}
We show that the conditional shape invariance symmetry can be used as a very powerful tool to calculate the eigenvalues of the mixed potential $V(r)= ar+br^{2}+\frac{c}{r}+\frac{l(l+1)}{r^{2}}$ for a restricted set of potential parameters. The energy for any state can be obtained algebraically, albeit for a severely restricted set of potential parameters. We also indicate that each member of the hierarchy of Hamiltonians is basically conditionally translational shape invariant. Comparison of analytically obtained results with numerical results is also presented. Our present methodology can be taken as an alternative treatment for the calculation of any higher order excited states of conditionally exactly solvable (CES) potentials.
\end{abstract}


%\begin{pacs}
 
%\end{pacs}

\end{frontmatter}

%\maketitle

\section{Introduction}

The connection between solvable potentials and supersymmetric quantum mechanics (SUSY QM) is an established fact \cite{1,2}. Gendenshtein showed that
whenever the special symmetry known as `shape-invariance symmetry' is satisfied by the supersymmteric partner potentials $V_{1}$ and $V_{2}$, the 
entire energy spectrum including eigenfunctions of the Hamiltonian can be calculated by purely algebraic means \cite{3}. This idea has been illustrated for the full set of exactly solvable potentials \cite{5}. However, rather more interesting scenario is the existence of conditionally exactly solvable(CES)
and quasi conditionally exactly solvable potentials (QCES). In the first case (CES), the full energy spectrum can be obtained only when the parameters of the
potential satisfy some particular condition. Whereas for the second case (QCES), a part of the eigen spectrum can be calculated only when a certain parametric condition is satisfied.
Some of these potentials which belong to the second class are Coulomb potential perturbed by terms of different powers of $r$ \cite{6,7,8}, anharmonic potential with quartic and sextic terms \cite{9,10} and the 
mixed potential \cite{11} which is the potential of our present study. In all these cases it has been demonstrated that the 
potential parameters can be chosen appropriately, such that exact analytic solution for the ground state of the potential can be obtained.
\\
In the present work, we revisit the mixed potential to establish the fact that a proper use of the conditional shape 
invariance symmetry is a very powerful tool to calculate analytically {\it the whole eigen spectrum} of the mixed potential. A number of 
investigations on this potential can be found in references \cite{12,13}, although none of them either established that the conditional shape 
invariance symmetry is responsible for occurrence of conditionally exact solution or calculated the whole spectrum. These earlier works 
have also not addressed the major issue of exploiting conditional shape invariance symmetry to get solution of higher excited states. 
Thus our present work will address two key features. First, we will explicitly show that this particular potential, which is a member of the CES potential family, basically inherits the conditional shape invariance symmetry. Second, we will utilize the CES symmetry in the hierarchy of Hamiltonians which leads to the very accurate results even for the excited states for judicious choices of potential parameters.\\
It is to be noted that numerical calculations of excited states for a given potential are more challenging, as they involve larger numerical errors due to the fact that the system becomes 
more extended as the excitation increases. Thus our present methodology has the potential to be presented as an accurate alternative treatment for the calculation of higher excited states of CES potentials.
For higher excited states of a given mixed potential, we will always look for the solution of the ground state of the next higher member
in the hierarchy of Hamiltonians. We will also show that the choice of parameters also plays a major role to get an accurate solution. We have compared
our analytic results with the numerical solution of the Schr\"{o}dinger equation and discussed both the demerits and efficacy of our methodology.\\
The plan of the paper is as follows. In Sec 2.1 we present a brief introduction to the SUSY QM language for general readers. In Sec 2.2 and Sec 2.3 we discuss the mixed potential extensively. In Sec 3 we present the comparison of numerical results with analytically calculated results. Sec 4 contains a summary followed by concluding remarks.

\section{Methodology}
\hspace*{.5cm}
\subsection{SUSY QM and shape-invariance}
To discuss the basics of SUSY formalism, we start with the SUSY Hamiltonian in one dimension, which is given by $ H= \left( \begin{array}{cc}
                                                               H_{1} & 0\\
                                                               0 & H_{2}\\
                                                              \end{array} \right)$ 
where the original Hamiltonian $H$ is factorized in two partner Hamiltonian $H_1$ and $H_2$.  $H_1$ and $H_2$ are defined as
\begin{equation}
 H_{1,2}= -\frac{d^{2}}{dx^{2}}+V_{1,2}(x),
\end{equation}
where the original potential $V(x)$ is also factorized in two supersymmetric partner potentials $V_{1,2}(x)$.
The partner potentials $V_{1,2}(x)$ are represented through the well-known Riccati equation
\begin{equation}
 V_{1,2}= W^{2}\mp W^{\prime}(x),
\end{equation}
where $ W(x)$ is referred as the superpotential and is related to the ground state wave function $\Psi_{0}(x)$ of $H_{1}$ through
\begin{equation}
 W(x)=-\frac{\Psi_{0}^{\prime}(x)}{\Psi_{0}(x)}.
\end{equation}
It is to be noted that energy levels (in the units of $ \frac{\hbar}{\sqrt{2m}}$) are  shifted to make the ground state energy of $H_1$ zero. Now in terms of generalized
annihilation and creation operator $A$ and $A^{\dagger}$, which are defined as  
\begin{eqnarray}
 A=\frac{d}{dx}+W(x) , \quad A^{\dagger}=-\frac{d}{dx}+W(x),   
\end{eqnarray}
the two partner Hamiltonians can be written as 
\begin{eqnarray}
 H_1= A^{\dagger}A, \quad  H_2= AA^{\dagger}. 
\end{eqnarray}
Using SUSY algebra, one can explicitly show the correspondence between energy levels of the two partner Hamiltonians. Let the eigenfunctions of $H_{1,2}$, which correspond 
to eigenvalues $E_n^{(1,2)}$ be $\Psi_{n}^{(1,2)}$. Then
\begin{eqnarray}
\begin{split}
 H_2(A\Psi_{n}^{(1)})= AA^{\dagger}(A\Psi_{n}^{(1)}) 
                     & = AH_1\Psi_{n}^{(1)} \\
                   & = E_n^{(1)}(A\Psi_{n}^{(1)}).
\end{split}
\end{eqnarray} 
Similarly
\begin{eqnarray}
\begin{split}
 H_1(A^{\dagger}\Psi_n^{(2)})= A^{\dagger}A(A^{\dagger}\Psi_n^{(2)})
                       &=A^{\dagger}H_2\Psi_n^{(2)}\\
                      & = E_n^{(2)}(A^{\dagger}\Psi_n^{(2)}).
\end{split}
\end{eqnarray}

Since $A\Psi_{0}^{(1)}=0$, ground state of $V_1(x)$ does not have a SUSY partner. So the pair of potential $V_{1,2}(x)$ have the same eigenspectra, only the partner state corresponding to the ground state of $V_{1}(x)$ is missing in the spectrum of $V_{2}(x)$. Thus, $E_{n+1}^{(1)}=E_{n}^{(2)},\quad E_0^{(1)}=0$.
\\ 
Gendenshtein pointed out that the shape-invariance condition is mathematically expressed \cite{3} as  
\begin{equation}
 V_{2}(x,a_{1})=V_{1}(x,a_{2})+R(a_{1}) ,\quad a_{2}= f(a_{1})
\end{equation}
where $a_{1}$ and $a_{2}$ are the two parameters appearing in the potential and $f$ is a prescription for getting $a_2$ from $a_1$ . To be similar in shape, the partner potentials must have same mathematical structure, but with altered parameters. It has been shown that this condition is an integrability condition \cite{4}
and applying SUSY algebra for a hierarchy of Hamiltonians along with this shape invariance condition, one obtains the full eigen-spectrum of $H_{1}$ as 
\begin{equation}
 E_{n}^{(1)}= \sum_{k=1}^{n} R(a_{k}).       %IS {(1)} OK?
\end{equation}
We will show that for CES potentials, the shape invariance criterion is satisfied only when a set of constraint conditions involving potential parameters 
is satisfied.
\subsection{Conditionally exact solution of mixed potential}
\hspace*{.5cm}
Algebraic method of solving a shape invariant potential by forming a hierarchy of Hamiltonians is well known and extensively used \cite{5}. Here, we are going to show its effective use in 
case of a CES potential also. The form of the potential of our present interest is 
\begin{equation}
 V(r)= ar+br^{2}+\frac{c}{r}+\frac{l(l+1)}{r^{2}}, \quad b > 0.
\end{equation}
So to calculate the ground state, we take 
\begin{equation}
 V_{1}(a,b,c,l,r)=V(r)-E_{0},
\end{equation}
where $E_{0}$ is the ground state energy of $V(r)$. Here by following the conventional formalism of SUSY we have subtracted $E_0$ to make the ground state energy of $V_1(r)$ zero and the Riccati equation is defined accordingly.  
We choose our superpotential ansatz as
\begin{equation}
 W(r)= Ar-\frac{B}{r}+D, \quad B > 0, A > 0.
\end{equation}
Now, $V_{1}(r)$ is related with $W(r)$ through the Riccati equation as
\begin{eqnarray}\label{12}
\begin{split}
 V_{1}(a,b,c,l,r)= V(r)-E_{0}= W^{2}(r)-W^{\prime}(r) \\
 = ar+br^{2}+\frac{c}{r}+\frac{l(l+1)}{r^{2}}-E_0,
 \end{split} 
\end{eqnarray}
where 
\begin{equation} \label{13}
%\begin{split}
  W^{2}(r)-W^{\prime}(r) = \frac{B^{2}-B}{r^{2}}-\frac{2BD}{r}+2ADr \\ 
   + A^{2}r^{2}+D^{2}-2AB-A. \\ 
%\end{split}
\end{equation}          
 Comparing the coefficients of each power of $r$ from eq. (\ref{12}) and eq. (\ref {13}), the unknown parameters are calculated and we also get a constraint condition as number of potential parameters are larger than the numbers of superpotential parameters.
The unknown parameters $ A, B, D$ of the superpotential are related with potential parameters through the following equations
\begin{eqnarray}
\begin{split}
   B^{2}-{B} & = l(l+1),\\
   A^{2} &  = b, \\
  -2BD & = c, \\
    2AD & =a,  \\
  D^{2}-2AB-A   & = -E_{0}.
 \end{split}
 \end{eqnarray}
The solutions of the above equations are 
\begin{eqnarray} \label{8}
A=\sqrt{b}, \qquad D=\frac{a}{2\sqrt{b}}, \qquad B=-\frac{\sqrt{b}c}{a}. \nonumber\\
\end{eqnarray}
Then the ground state energy of $V(r)$ is given by
\begin{equation}
 E_{0}=-\frac{a^{2}}{4b}-\frac{2bc}{a}+\sqrt{b},
\end{equation}
subject to the parametric constraint condition
\begin{equation}\label{1}
 l(l+1)= \frac{c^{2}b}{a^{2}}+\frac{c\sqrt{b}}{a}.
\end{equation}
It is evident from the above discussion that not all the potential parameters can be chosen freely as the constraint condition regarding potential parameters has to be satisfied. Now, a feature of conditional shape invariant potential is that the number of potential parameters are generally larger than the number of superpotential parameters. So there is a constraint condition relating potential parameters (eq. \ref{1}). We have chosen $l$, $b$, $c$ as free parameters and then $a$ is calculated from the constraint condition 
and calculated values of $a$ are presented in Table 2 for different sets of potential parameters. \\ 
Now to check the shape invariance property of $V_{1}(r)$ we construct the SUSY partner potential $V_{2}(r)$ as
\begin{eqnarray}\label{3}
\begin{split}
 V_{2}(a,b,c,l_{1}, r) & =W^{2}(r)+W^{\prime}(r)\\
  & = \frac{B^{2}+B}{r^{2}}-\frac{2BD}{r}+2ADr\\
   & +A^{2}r^{2}+D^{2}-2AB+A \\
    &  =ar+br^{2}+\frac{c}{r}+\frac{l_{1}(l_{1}+1)}{r^2}+R.
\end{split}
\end{eqnarray}
Again the functional relation between parameters of $V_2(r)$ and $W(r)$ can be calculated by comparing the coefficients of different powers of $r$. The calculation reveals that $b$, $c$ and $a$ remain same as their functional dependences with respective superpotential parameters remain unchanged, although $l$ has been changed to $l_1$ while $l_1$ is determined by
\begin{equation}\label{2}
l_1(l_1+1)= B^{2}+B.
\end{equation} 
From the structure of eq. (\ref{2}), it can be shown that the new parameter ($l_1$) is related with the old parameter ($l$) by translation as
\begin{eqnarray}
\begin{split}
 l_{1}(l_{1}+1) &=B^{2}+B \\
             &  =\frac{c^{2}b}{a^{2}}-\frac{c\sqrt{b}}{a} \\
               & =l(l+1)-2\frac{c\sqrt{b}}{a} \\
             &  =(l+1)(l+2), 
\end{split}
\end{eqnarray}
Thus $l$ is translated to $l+1$ during the passage of $V_1(r) \rightarrow V_2(r)$
and $R$ is determined through the relation  
\begin{equation}\label{4}
 R= D^{2}-2AB+A.
\end{equation} \\
 Next, we carry forward this technique to the calculation of excited states of $V(r)$. As $V_1(r)$ and $V_2(r)$ are the two partner potentials, they maintain the energy degeneracy relation and ground state of $V_2(r)$ is nothing but the first excited states of $V_1(r)$. Now to calculate the ground state of $V_2(r)$, we start from $V_2(r)$ as the new starting potential and define  $V^{\prime}(r)= V_2(r)$. If $E^{\prime}_{0}$ is the ground state energy of $V^{\prime}(r)$, we shift the energy scale again by $E^{\prime}_{0}$, so that its ground state is at zero energy 
\begin{equation}
\begin{split}
 V_1^{\prime}(r) &=V^{\prime}(r)-E^{\prime}_{0}.
\end{split}
\end{equation}
 We start with a similar anstaz for the superpotential of $V^{\prime}(r)$ as
\begin{equation}
\hat{W}(r)= A^{\prime}r-\frac{B^{\prime}}{r}+D^{\prime}, \quad B^{\prime} >0, A^{\prime} >0  
\end{equation}
 Repeating the same procedure as before, we find that the superpotential parameters ($A^{\prime}, B^{\prime}, D^{\prime}$) have similar relationships with parameters of the corresponding potential, as in the previous case.  Note that we have chosen $l, b, c$ as independent parameters and $a$ is calculated through the constraint condition. As during the passage of calculating first member in the hierarchy to the next one $l$ is translated to $l+1$, so the calculated value of $a$ will be different from it's previous value. We call it $a_1$. Thus $A^{\prime}, B^{\prime}, D^{\prime}$ are now related with $a_1, b, c$ by 
\begin{eqnarray}\label{9}
A^{\prime}=\sqrt{b}=A, \qquad D^{\prime}=\frac{a_{1}}{2\sqrt{b}}, \qquad B^{\prime}=-\frac{\sqrt{b}c}{a_{1}}. \nonumber\\
\end{eqnarray} 
So the constraint condition for ground state is given by  
\begin{equation}\label{7}
  l_{1}(l_{1}+1)=\frac{c^{2}b}{a_{1}^{2}}+\frac{c\sqrt{b}}{a_{1}}
\end{equation}
where $l_1=l+1$. 
Then the ground state energy of $ V^{\prime}(r)$ is 
\begin{eqnarray}
\begin{split}
 E_{0}^{\prime} & =\frac{a^{2}}{4b}+\frac{2bc}{a}+\sqrt{b}+(-\frac{a_{1}^{2}}{4b}-\frac{2bc}{a_{1}}+\sqrt{b}) \\
          & = \frac{a^{2}}{4b}-\frac{a_{1}^{2}}{4b}+\frac{2bc}{a}-\frac{2bc}{a_{1}}+2\sqrt{b}.
 \end{split}
\end{eqnarray}
 We have calculated the ground state of $V^{\prime}(r)$ which is actually the first excited state of $V_{1}(r)$. Now as ground state energy of $V_{1}(r)$ has been made zero by shifting energy scale,  so the calculated ground state energy of $V^{\prime}(r)$ have to be properly shifted by adding the ground state energy of $V(r)$. Therefore, in proper energy scale, the ground state energy of $V^{\prime}(r)$ is 
\begin{equation}
 E_{0}^{V^{\prime}}=E_{0}^{\prime}+E_{0}=-\frac{a_{1}^{2}}{4b}-\frac{2bc}{a_{1}}+3\sqrt{b}.
\end{equation} \
Thus from the above discussion, we claim that the given mixed potential is shape invariant but this shape invariance criterion holds only when the potential parameters satisfy a particular constraint condition. So we designate it as conditional shape invariance symmetry. However it is desired to see the equivalent relation through the Riccati equation which is represented as
\begin{eqnarray}\label{10}
\begin{split}
W^{2}(r;A,B,D)+ \frac{dW(r;A,B,D)}{dr} \\
 = W^{2}(r;A^{\prime},B^{\prime},D^{\prime})- \frac{d(W(r;A^{\prime},B^{\prime},D^{\prime})}{dr}+R.
\end{split}
\end{eqnarray}
For the traditional exactly shape invariant potential, $R$ is exactly the ground state energy as one-to-one correspondence between the potential parameters and those of superpotential parameters holds good. However direct satisfaction through the algebraic relation as given by eq. (\ref{10}) is not possible as our potential is conditionally shape invariant and there is no clear one-to-one correspondence between potential parameters and superpotential parameters. Because of the presence of constraint condition, one parameter is always dependent on the particular choice of potential. Nonetheless, if the left side of the eq. (\ref{10}) is calculated by using eq. (\ref{3}) together with the value of $A, B, D$ from eq. (\ref{8}) and right side of the eq. (\ref{10}) is calculated by using eq. (\ref{13}) together with the value of $A^{\prime}, B^{\prime}, D^{\prime}$ from
 eq. (\ref{9}) and  $R$ from eq. (\ref{4}), then the algebraic relation presented by eq. (\ref{10}) is exactly satisfied. One can easily verify it by putting the value of $b, c, 
l$ and $a$ taken from Table 2 from the presented value of potential parameters in section 3.   \\

Now following the same procedure we can construct the partner of $V^{\prime}$ (say $V^{\prime \prime}$) and partner of $V^{\prime \prime}$ (say $V^{\prime \prime \prime}$) and partner of $V^{\prime \prime \prime}$ (say $V^{\prime \prime \prime \prime}$). Thus $V(r)$, $V^{\prime}(r)$, $V^{\prime \prime}(r)$, $V^{\prime \prime \prime}(r)$, $V^{\prime \prime \prime \prime}(r)$ are
basically the members of the hierarchy of Hamiltonians which are constructed by basic SUSY mechanism in each step. In construction of this hierarchy we utilise the same form of superpotential with different parameters and relate the potential parameters with those of corresponding superpotential parameters, which in turn give us the constraint conditions regarding potential parameters along with the ground state energies.  However, during transition from $V(r)\rightarrow V^{\prime}(r)$, $V^{\prime}(r) \rightarrow V^{\prime \prime}(r)$,
$V^{\prime \prime}(r) \rightarrow V^{\prime \prime \prime}(r)$, $V^{\prime \prime \prime}(r) \rightarrow V^{\prime \prime \prime \prime}(r)$, etc., we notice that a new constraint condition has to be satisfied in each step, since the parameter $l$ is translated as $ l\rightarrow (l+1) \rightarrow (l+2) \rightarrow (l+3) \rightarrow (l+4)$, etc. respectively.
 We observe that the pair of partner potentials ($V_1$, $V^{\prime}$){;}($V^{\prime}$, $V^{\prime \prime}$){;}($V^{\prime \prime}$, $V^{\prime \prime \prime}$){;}($V^{\prime \prime \prime}$, $V^{\prime \prime \prime \prime}$) are
shape invariant  but shape invariance is satisfied only when potential parameters satisfy the particular constraint condition. Thus the first and preliminary finding of our investigation is that the mixed potential is conditionally exactly solvable because it satisfies shape invariance subjected to a particular constraint condition involving potential parameters.\\[.1cm]
 The second motivation of our present work is to extend the concept of conditional shape invariance for excited states. Here we take the opportunity to extend our 
calculation to the excited states of stationary Hamiltonian and to see how accurately that can be calculated with the ground state of different members of the hierarchy. For truly shape invariant potentials this is a trivial task as the 
ground state of $nth$ member of the hierarchy basically corresponds to $(n-1)th$ excited state of the original Hamiltonian. But in case of mixed potential, due to additional conditional shape invariance criterion this is no more a trivial task. Together with this, as all of the potential parameters are not freely chosen, the accuracy of SUSY result will highly depend on the choice of parameters.\\
In the following section we try to exhibit the degeneracy property of members of the  hierarchy of potentials and show that
\begin{eqnarray}
 E_1^{V_1}=E_0^{V^{\prime}} \\
 E_2^{V_1}=E_1^{V^{\prime}}=E_0^{V^{\prime \prime}} \\
 E_3^{V_1}=E_2^{V^{\prime}}=E_1^{V^{\prime \prime}}=E_0^{V^{\prime \prime \prime}}\\
 E_4^{V_1}=E_3^{V^{\prime}}=E_2^{V^{\prime \prime}}=E_1^{V^{\prime \prime \prime}}=E_0^{V^{\prime \prime \prime \prime}}.
\end{eqnarray} 
\subsection{General form of energy and constraint condition}
In this section we derive a general formula for calculation of the parameter $a$, as well as the energy of any excited state in terms of independent parameters $c$, $b$ and $l$.
For the first excited state if we write the constraint condition as a quadratic equation in $a$ (renaming as $a_1$), then 
\begin{equation}
 l_1(l_1+1)a_1^2-c\sqrt{b}a_1-c^2b=0,\quad l_1=l+1
\end{equation}
 Solving this equation, we get two solutions for $a_1$ 
 \begin{eqnarray}\label{5}
  a^{(+)}_1= \frac{c\sqrt{b}}{l+1} \\
  a^{(-)}_1= - \frac{c\sqrt{b}}{l+2} 
  \end{eqnarray}
The second solution for $a_1$ has been chosen. It is evident from the construction of hierarchy of Hamiltonians that for different members of the hierarchy, constraint conditions are also different, as parameters are translated. Of the two possible solutions for $a_1$, only $a_1^{(-)}$ is parametrically translated properly from $a$, as we have taken $a=-\frac{c\sqrt{b}}{l+1}$ for the ground state. So we have chosen the second solution as only this solution incorporates the translational change of $l$. As $b$ and $c$ are independent parameters, they are constants for all $V_n(r)$. Let for $V_k(r)$, $ a = a_k$, $l= l_k$. Then for $V_{k+1}(r)$ we have $a_1=a_{k+1}$ and $l_{k+1}=l_k+1$. 
Generalizing eq. (34), we obtain the exression of $a_k$ for any higher order excited state, which is given by
\begin{equation}
  a_k= -\frac{c\sqrt{b}}{l_k+2},\quad k =1,2,3..
\end{equation}
where       %$a_1=a$,\ 
$l_1=l$. Now in terms of ground state angular momentum $l$,\ $a_k$ can be written as
\begin{equation}\label{6}
  a_k= -\frac{c\sqrt{b}}{l+k+1},\quad  k =0,1,2,3..
\end{equation}
It is to be noted that from eq. (\ref{6}), the parameter $a$ can be calculated for the ground state as well as for any excited state.\\
Hence the ground state energy in terms of independent parameters is 
\begin{equation}
 E_0= -\frac{c^2}{4(l+1)^2}+2\sqrt{b}(l+1)+\sqrt{b}.
\end{equation}
Now the ground state energy of $V_2(r)$ in properly shifted scale, which is the first excited state energy of $V(r)$, is given by
\begin{eqnarray}
\begin{split}
 E_{0}^{V_2} & =E_{0}^{\prime}+E_{0} \\
          &  =-\frac{c^2}{4(l_1+1)^2}+2\sqrt{b}(l_1+1)+3\sqrt{b}.
 \end{split}
\end{eqnarray}
Generalizing, ground state energy of $V_k(r)$, which is the $(k-1)$th excited state of $V(r)$, is 
\begin{equation}
 E_k= -\frac{c^2}{4(l+k+1)^2}+2\sqrt{b}(l+1)+(4k+1)\sqrt{b}, \quad k=0,1,2,..
\end{equation}
As the expression  is written in terms of independent parameters $b$, $c$, which are constants, any excited state energy of $V(r)$ can be easily calculated.  So all the excited states of $V_k(r)$ can be obtained exactly and analytically, provided $a_k$ satisfies the constraint condition which highlights the fact that in this case exact solution is a manifestation of conditionally shape invariance symmetry. 
\section{Results and Discussion}
\hspace*{.5cm}
For the first choice (denoted by I) we take $l=1,\ c=0.01,\ b=0.5 $. And the parameter $a$ is equal to $(-3.53\times 10^{-3})$ which is calculated from the conditional shape invariance criterion. The summarized results for energy calculation are presented in Table 1.  All the calculated energies are in the original scale.
\begin{table}[h!]
\centering
\caption{Comparison of energies of ground and excited states of $V(r)$ obtained from SUSY formalism with the numerical results . Choice of parameters corresponds to set I: $l$=1, $b$=0.5, $c$=0.01}
\label{my-label} 
\begin{tabular}{|c|c|c|c|c|c|}
\hline
V              & V$^{\prime}$   & V$^{\prime\prime}$  & V$^{\prime\prime\prime}$ & V$^{\prime\prime\prime\prime}$    &$\varDelta$E    \\ \hline 
E$^{S}_{0}$=3.53                & $-$                      & $-$                  &  $-$                  & $-$    &  exact     \\ \cline{1-1}
E$^{R}_{0}$=3.53 
                                    &                       &                       &                         &              &       \\ \hline
E$^{R}_{1}$=6.36        & E$^{S}_{0}$=6.36           & $-$                  & $-$                     & $-$         &   0         \\ \cline{2-2} 
                          & E$^{R}_{0}$= 6.36          &                       &                         &               &             \\ \hline
E$_2^{R}$=9.19          &E$_1^{R}$=9.19               & E$^{S}_{0}$= 9.19  & $-$                  & $-$          &  0             \\ \cline{3-3} 
                              &                         & E$^{R}_{0}$=9.19     &            &              &           \\ \hline
E$_3^{R}$=12.02          & E$_2^{R}$=12.02       & E$_1^{R}$=12.01     & E$^{S}_{0}$=12.02    & $-$        & 0.0           \\ \cline{4-4} 
                              &                    &                    & E$^{R}_{0}$=12.01    &        &               \\ \hline
E$_4^{R}$=14.92         & E$_3^{R}$=14.87    & E$_2^{R}$=14.86    & E$_1^{R}$=14.85       & E$^{S}_{0}$=14.85 & 0.07   \\ \cline{5-5} 
                               &                    &                       &               & E$^{R}_{0}$= 14.84  & \\ \hline
\end{tabular}
\end{table}

For comparison, we also present numerical solution of Schr\"{o}dinger equation by the RKGS method for all states. $E_{0}^{SUSY}$ is the exact SUSY result and $E_{n}^{RKGS}$ $(n=0,1,2..)$ are the RKGS results for the $n^{th}$ state of a given potential. In the table, $E^S$ represents energy value calculated from SUSY formalism, $E^R$ represents energy value calculated using RKGS method.
The accuracy of the calculation of higher excited states is defined as $ \varDelta E=E^{n}_{RKGS}-E^{n+1}_{0,SUSY}$,\  [$n$=1,2,3,4] where $E_{RKGS}^n$ is the numerical solution of Schr\"{o}dinger equation
for $n$th excited state of $V$ and $E^{n+1}_{0,SUSY}$ is the SUSY energy for the ground state of $(n+1)$th Hamiltonian in the hierarchy and is calculated from the corresponding analytic expression. \\
In the first column we present the ground state energy as well as the excited state energy of $V(r)$. It is to be noted that the ground state of each member of the hierarchy is exactly reproduced in SUSY as expected. It is also very gratifying to see that
the first and higher excited states are very accurately calculated using basic SUSY mechanism along with conditional shape invariance condition.
The very close agreement between SUSY and RKGS for the excited states lies on the fact that calculated values of $a$ do not change significantly during the passage from one member to the next member of the hierarchy as value of $a$ itself is very small.\\
In Table 2, we present the value of $-a$ calculated for different partners (we have given the numerical values with three significant digits).
\begin{table}[h!]
\centering
\caption{ Calculated values of the dependent parameter $-a$ for the two different sets}
\label{my-label}
\begin{tabular}{|c|c|c|c|c|c|}
\hline
 No.  & V    & V$^{\prime}$   & V$^{\prime\prime}$   & V$^{\prime\prime\prime}$ & V$^{\prime\prime\prime\prime}$      \\ \hline
 I  & $3.53\times 10^{-3}$  & $2.35\times 10^{-3}$  & $1.76\times 10^{-3}$  & $1.41\times 10^{-3}$  & $1.17\times 10^{-3}$ \\ \hline
 II & 0.866    & 0.577        & 0.433         & 0.346     & 0.288  \\  \hline
\end{tabular}
\end{table}
\\

For the second choice (denoted by II) we take  $l=1,\ b=3.0,\ c=1.0$ and the results are summarized in Table 3. The corresponding value of $-a$ for different members of the hierarchy has been presented in Table 2.
Here we observe that the ground state energy is reproduced exactly for $V$ but higher excited states are slightly above the SUSY results. The $\varDelta E$ value gradually increases with the number of excited states. From Table 2, it is quite easy to see that the calculated values of the dependent parameter $a$ obtained from the translational but conditional shape invariance during the passage of each member of the hierarchy does not change significantly for the first choice. However in the second choice the change in $a$ is noticeable. So in the second case the effect of change in $a$ for differnt members of Hamiltonian hierarchy is more pronounced in the potential.  
\begin{table}[h!]
\centering
\caption{Comparison of energies of ground and excited states of $V(r)$ obtained from SUSY formalism with the numerical results. Choice of parameters corresponds to Set II: $l$= 1, $b$= 3.0, $c$= 1.0}
\label{my-label}
\begin{tabular}{|c|c|c|c|c|c|}
\hline
V                & V$^{\prime}$ & V$^{\prime\prime}$ & V$^{\prime\prime\prime}$ & V$^{\prime\prime\prime\prime}$     &$\varDelta$E    \\ \hline
 E$^{S}_{0}$=8.59                & $-$                 &  $-$                 & $-$        & $-$   & exact \\ \cline{1-1}
E$^{R}_{0}$=8.59
                                    &                       &                  &                &                 &   \\ \hline
E$^{R}_{1}$=15.59     & E$^{S}_{0}$=15.56           & $-$               & $-$            & $-$   & 0.04\\ \cline{2-2}
                         & E$^{R}$= 15.56         &                       &                       &         &    \\ \hline
E$_2^{R}$=22.58      & E$_1^{R}$= 22.29  & E$^{S}_{0}$= 22.49         & $-$          & $-$    & 0.09 \\ \cline{3-3}
                               &          & E$^{R}_{0}$=22.32            &         &           & \\ \hline
E$_3^{R}$=29.51      & E$_2^{R}$=29.01  & E$_1^{R}$=29.12     & E$^{S}_{0}$=29.42  & $-$  &0.09 \\ \cline{4-4} 
                                &                    &          & E$_{0}^{R}$=29.39   &          &   \\ \hline
E$_4^{R}$=36.45       & E$_3^{R}$=35.80 & E$_2^{R}$=35.98     & E$_1^{R}$=36.03 & E$^{S}_{0}$=36.35   & 0.10  \\ \cline{5-5} 
                              &                    &                       &     & E$^{R}_{0}$= 36.09    &   \\ \hline
\end{tabular}
\end{table}

For further clarification, in Fig.1 and in Fig.2 we plot $V$, $V^{\prime}$, $V^{\prime\prime}$, $V^{\prime\prime\prime}$, $V^{\prime\prime\prime\prime}$ together.
\begin{figure}[h]
 \vspace{-10pt}
\begin{center}
%\includegraphics{width=0.5\textwidth, inner}
%\begin{tabular}{cc}
%\hspace{-3.3mm}
%\rotatebox{0}{\epsfxsize=7.8cm\epsfbox{fig1.eps}}
\resizebox{60mm}{!}{\includegraphics[angle=0]{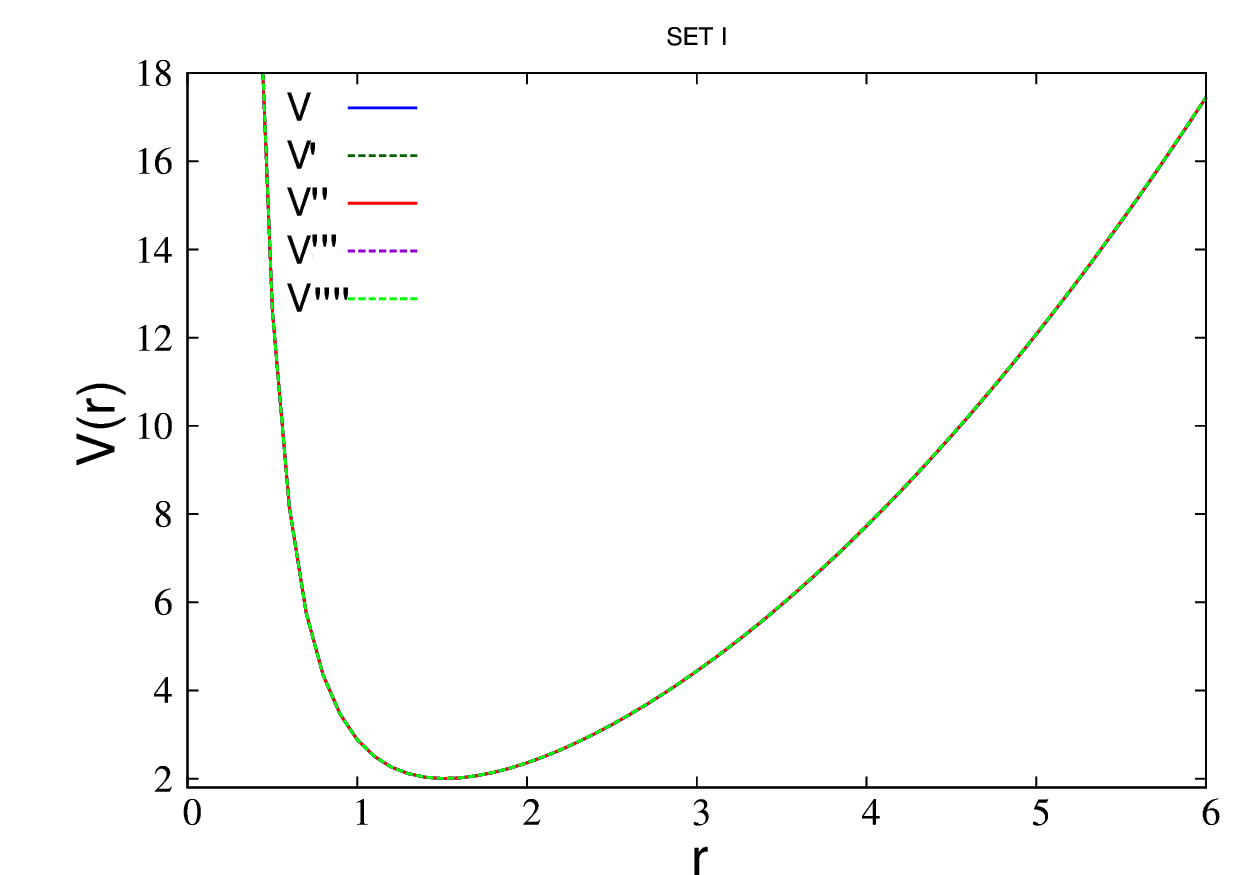}} 

\caption{Plot of several partner potentials $(V, V^{\prime}, V^{\prime\prime}, V^{\prime\prime\prime}$ and $V^{\prime\prime\prime\prime})$ in the appropriate scale for the first choice of potential parameters.}
\end{center}
\end{figure}
For the first choice all the four members basically overlap and it is very hard to distinguish them. Although $a$ is calculated from constraint condition in each step, the variation in $a$ is insignificant as value of $a$ is very small.\\
But for the second choice, the change in calculated values of $a$ is significant. During passage of $V\rightarrow V^{\prime}\rightarrow V^{\prime\prime}\rightarrow V^{\prime\prime\prime}\rightarrow V^{\prime\prime\prime\prime}$, the value of $a$ becomes less negative which makes
the potential less negative.
\begin{figure}[h]
 \vspace{-10pt}
\begin{center}
%\hspace{-3.3mm}
%\rotatebox{0}{\epsfxsize=7.8cm\epsfbox{fig2.eps}}
\resizebox{60mm}{!}{\includegraphics[angle=0]{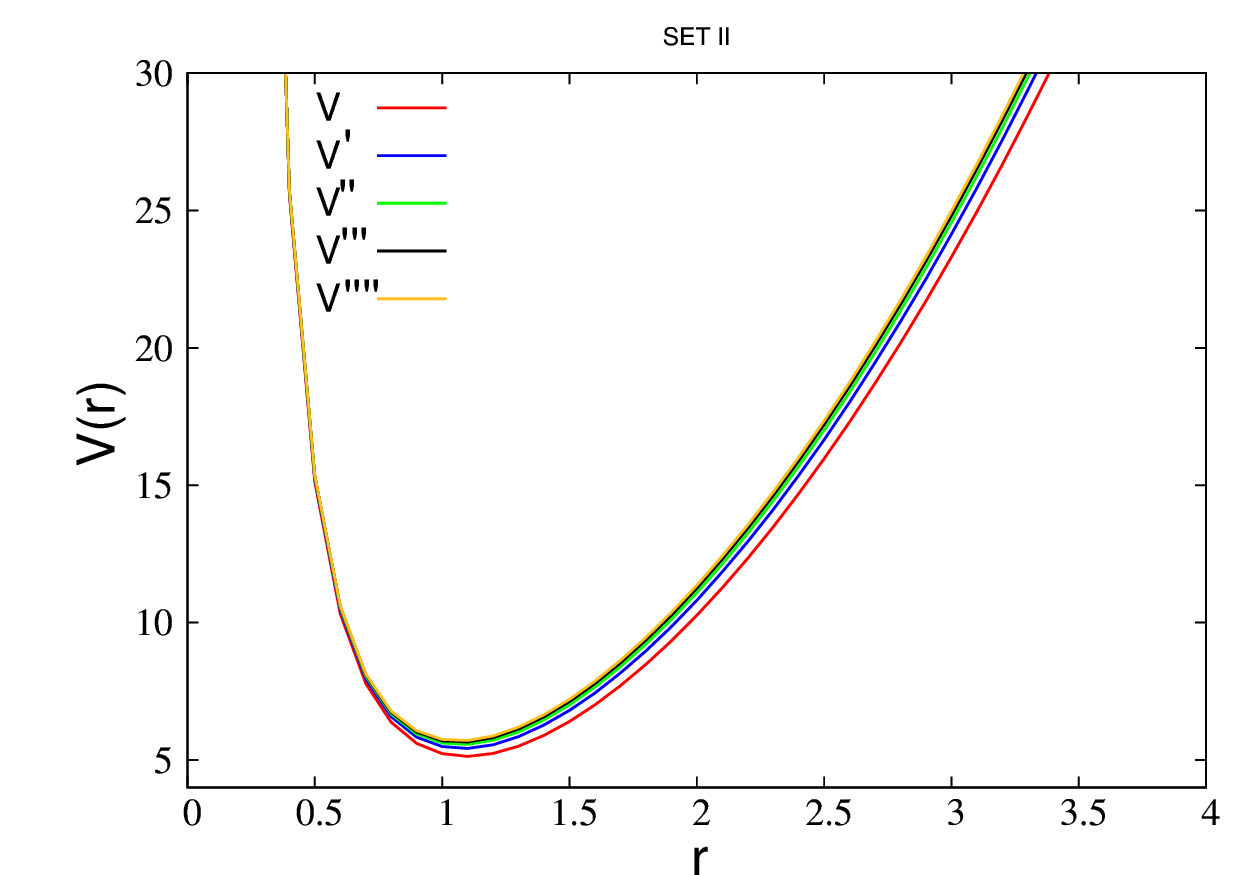}} 
\caption{Plot of several partner potentials $(V, V^{\prime}, V^{\prime\prime}, V^{\prime\prime\prime}$ and $ V^{\prime\prime\prime\prime})$ in the appropriate scale for the second choice of potential parameters.}
\end{center}
\end{figure}

Thus the SUSY values for the excited states of $V(r)$ are lower than the numerical values. The effect is more significant for higher order excited states.\\
One may be curious to know what will be the result for some other choices of $a$. For some other choice of $a$ where the difference of $a$ is significant for any two partner potential, the partner potentials will be more distinct. Also for such a set of parameters, although the ground state energy will always be exact, analytically and numerically calculated energy values may differ significantly for the excited states. It is to be noted that in this paper, our main motivation is to depict the efficacy of SUSY methodology and conditional shape invariance symmetry. So we have deliberately kept those choices where $ \varDelta E$ is very small.   

\section{Conclusion} 

In the present manuscript we have discussed the conditional shape invariance symmetry of a mixed potential. We first show that conditional shape invariance is the required symmetry to have the conditional exact solution of this potential. Our superpotential ansatz is a very powerful technique where no initial input is necessary. Utilizing the Riccati equation we relate the unknown parameters of the superpotential in terms of the potential parameters. Next we develop the hierarchy of Hamiltonians and highlight the fact that during the passage from one member to the next of the hierarchy of  Hamiltonians the conditional shape invariance symmetry is necessary. Finally in the second part of our work we have extensively used the conditional shape invariance property as a very powerful tool to get the excited states of the original potential for a restricted set of potential parameters. In each case we look for the exact algebraic result for the ground state of the SUSY partners, then SUSY results are further 
compared with the numerical solution 
of Schr\"odinger equation. The open question in this discussion remains wheather it is possible to extend our present methodology for the other choice of CES potentials where more than one constraint condition has to be satisfied. One such potential is the generalized polynomial potentials \cite{14}.


\begin{thebibliography}{References}
\bibitem{1} E. Witten, Nucl. Phys. B 188 (1981) 513
\bibitem{2} R. Dutt, A. khare, U. Sukhatme Am. Jour Phys. 56 (1988) 163
\bibitem{3} L. E. Gendenshtein, JETP Lett. 38 (1983) 356 
\bibitem{4} L. Infeld and T.E. Hull, Rev. Mod. Phys. 23, 21-68 (1951)
\bibitem{5} F. Cooper, A. Khare, U. S. Sukhatme, Phys. Rep. 251 267(1995) and reference therein.
\bibitem{6} A. de Souza Dutra, Phys.Review A 47 (1993) 4
\bibitem{7} A. de Souza Dutra, Phys. Lett. A 131 (1988) 319
\bibitem{8} G. P. Flessas, J. Phys. A 15, L1 (1982)
\bibitem{9} B. Chakrabarti  J. Phys. A: Math. Theor. 41 (2008) 405301
\bibitem{10} R. Roychoudhury, Y. P. Varshni, M. Sengupta, Phys. Rev. A42 (1990) 184
\bibitem{11} R. N. Chaudhuri and M.Mondal Phys. Review A 52 (1993) 3
\bibitem{12} V. Gupta, A. Khare, Phys. Lett. 70B, 313 (1977)
\bibitem{13} R. Roychoudhury, P. Roy, M. Znojil, Geza Lvai J. Math. Phys. 42 (1996) 2001
\bibitem{14} R. Adhikari, R. Dutt, Y.P. Varshni, Phys. Lett. A 141, 1 (1989)






\end{thebibliography}
\end{document}